\documentclass[11pt,twoside]{article}

\usepackage{asp2006}
\usepackage{epsf}
\usepackage{psfig}
\usepackage{lscape}

\begin{document}

\title{Physics of pulsar radio emission outside of the main
pulse}
\author{S. A. Petrova}
\affil{Institute of Radio Astronomy, Chervonopraporna Str., 4, Kharkov 61002, Ukraine}    

\begin{abstract} 
We for the first time propose a physical model of the precursor (PR) and interpulse (IP) components of the radio pulsar profiles. It is based on propagation effects in the secondary plasma flow of a pulsar. The components are suggested to
result from the induced scattering of the main pulse (MP) into background.
The induced scattering appears efficient enough to transfer a significant
part of the MP energy to the background radiation. In the regimes of
superstrong and moderately strong magnetic field, the scattered components are approximately parallel and antiparallel to the velocity of
the scattering particles and can be identified with the PR and
IP, respectively. The spectral evolution, polarization properties, and fluctuation behaviour of the scattered components are examined and compared with the observational results. The perspectives of the complex profile studies are outlined as well.
\end{abstract}

\section{Introduction}
The profiles of some radio pulsars contain the additional components
outside of the MP. The PR precedes the MP by a few tens degrees, whereas the IP lags the MP by about a half of the pulsar period. The components show peculiar spectral,
polarization and fluctuation properties and are usually interpreted in terms of several geometrical models. The peculiar locations of the components on the pulse profile are generally explained as the consequence of their generation in widely separated regions of the pulsar magnetosphere in the case of an especial geometry of the pulsar (the magnetic and rotational axes should be either nearly orthogonal or nearly aligned). However a bulk of observational evidence indicates that the PR and IP are physically related with each other and with the MP. Furthermore, recent observations of PSR B1702-19 have revealed a stiff connection between the subpulse modulation in the IP and MP \citep*{p2-w07}, which directly conflicts with the assumption of an independent generation of the components in different regions of the magnetosphere.
We for the first time propose a physical model of the PR and IP based on propagation effects in pulsar magnetosphere \citep[see also][]{p2-p08a,p2-p08b,p2-p08c}.

\section{Induced scattering of the MP into background}

The brightness temperatures of pulsar radio emission are so high that the induced scattering off the secondary plasma particles may be significant.	The pulsar radio emission is known to be highly directional: at any point of the emission cone it is concentrated into a narrow beam of the opening angle $\approx 1/\gamma$, where $\gamma$ is the particle Lorentz-factor,	and the beam propagates at the angle $1/\gamma\ll\theta\la 1$ to the magnetic field.	Outside of the beam, there may be some background photons, e.g. because of the spontaneous scattering from the beam.	In the course of induced scattering of the beam photons into the background, the background intensity grows exponentially, $I_{\rm bg}\propto\exp(\Gamma)$, with the scattering efficiency $\Gamma$ depending on the orientation of the background photons.	As a result, there arises a strong narrow component in the direction corresponding to the maximum scattering probability.	Given that the scattering is efficient, a significant part of the beam energy can be transferred to the background.	As the initial background intensities are extremely small, $I_{\rm bg}^{(0)}/I_{\rm beam}^{(0)}=10^{-8}-10^{-12}$, the induced scattering becomes efficient at $\Gamma=20-30$. Such values can indeed be met in pulsars.

The presence of a strong magnetic field affects the scattering process, modifying both the scattering cross-section and the geometry of recoil. This occurs on condition that $\omega^\prime\ll\omega_G$, where $\omega^\prime$ is the radio frequency in the rest frame of the scattering particles and $\omega_G\equiv eB/mc$ is the electron gyrofrequency. In contrast to the non-magnetic case, in the course of the magnetized induced scattering the intensity is redistributed in such a way that the total intensity of the beam and the background remains approximately constant, $I_{\rm beam}+I_{\rm bg}\equiv I_{\rm beam}^{(0)}+I_{\rm bg}^{(0)}$. The background intensity increases exponentially with $\Gamma$ until $I_{\rm bg}\sim I_{\rm beam}^{(0)}$. At still larger $\Gamma$ it stops growing and enters the stage of saturation, whereas the beam intensity drops significantly.
There are two regimes of the magnetized induced scattering, which hold in pulsar magnetosphere.

In a superstrong magnetic field,	the perturbed motion of the scattering particle in the fields of the incident wave is confined to the magnetic field line, causing a number of peculiarities in the scattering process. This is the so-called longitudinal scattering. In this case, only the waves of one polarization type (the ordinary waves, with the electric vector in the plane of the ambient magnetic field) are involved in the scattering. The scattered component arising in the longitudinal regime is directed approximately along the ambient magnetic field. Because of rotational aberration it appears on the pulse profile about $r_s/2r_L$ in advance of the MP (here $r_s$ is the altitude of the scattering region and $r_L$ is the light cylinder radius, $r_s =(0.1-1)r_L$) and can be identified with the PR.

In a moderately strong magnetic field,	the components of the excited motion of the particle, which are perpendicular to the field line, play a substantial role. In this so-called transverse regime,	the scattered radiation is a mixture of the two types of waves, with the ordinary and extraordinary polarization.	The scattered component is antiparallel to the longitudinal velocity of the scattering particles and can be recognized as the IP. The induced scattering switches between the longitudinal and transverse regimes on condition that $\omega^\prime/\omega_G=\gamma^{-1/2}$ \citep{p2-p08b}.

\section{Observational consequences}

\subsection{Spectral properties}

For both the PR and IP the altitudes of the scattering regions are related to the radius of cyclotron resonance and do not change with frequency substantially. Correspondingly,	the component locations on the pulse profile should be almost independent of frequency. This is really typical of the components outside of the MP.

At lower frequencies the efficiency of induced scattering is larger. Therefore the spectra of the scattered components should be steeper than the MP spectrum, in agreement with the observed trend. Besides that,	the MP spectrum is believed to flatten as a result of the scattering.

\subsection{Polarization of the scattered components}

The radiation resulting from the longitudinal scattering has only the component of the ordinary polarization, so that the PR is completely linearly polarized in the plane of the external magnetic field.	Complete linear polarization is indeed a characteristic feature of the observed PRs. In the case of transverse scattering, both types of the polarization, the ordinary and extraordinary ones, are present in the scattered radiation. However the scattering efficiency in different polarization channels differs substantially. Then the scattered radiation should be dominated by one type of polarization, and the IP is expected to be strongly polarized. Note that the degree of the IP polarization should depend on the scattering efficiency, so that the strongest IPs should have the strongest polarization, in agreement with the observed trend.

For the scattered components the position angle of linear polarization is determined by the orientation of the ambient magnetic field in the scattering region and therefore differs from the position angle of the MP, which is determined by the magnetic field orientation in the emission region.	In the outer magnetosphere, where the scattering takes place, the radio beam spans only a small part of the cross-section of the open field line tube, and the magnetic field orientation is almost unchanged across the scattering region.	Therefore the position angle of the scattered components should remain almost constant, in agreement with the observations.

\subsection{Intensity statistics of the scattered components}

In the rest frame of the scattering particles, the photon frequency is almost unchanged during the induced scattering; in the observer frame, the frequencies of the incident and scattered radiation, $\omega$ and $\omega_1$, are related as $\omega (1-\beta\cos\theta)=\omega_1(1-\beta\cos\theta_1)$, where $\beta\equiv (1-\gamma^{-2})^{1/2}$ is the velocity of the scattering particles in units of $c$, $\theta$ and $\theta_1$ are the wavevector tilts of the incident and scattered waves to the ambient magnetic field.	For the PR $\theta_1\approx 1/\gamma$ and $\omega_1\approx\omega\theta^2\gamma^2\gg\omega$, so that this component is fed by the MP radiation of lower frequencies.	As at lower frequencies the intensity is higher, the MP scattering into the PR leads to the increase of the profile intensity at a fixed frequency.	This is in line with the observational data on the Vela pulsar \citep{p2-kd83} and PSR B1822-09 \citep{p2-f82}, which testify to the greater role of the PR in stronger pulses.

In the case of IP $\theta_1\approx\pi$ and $\omega_1\approx\omega\theta^2<\omega$, so that the IP is weaker than the MP of the same frequency, if only the latter is not suppressed by the scattering substantially.	According to \citet{p2-f82}, the IP of PSR B1822-09 is indeed pronounced in the weaker pulses. Furthermore, in a number of pulsars the IP intensities are only a few per cent of those of the MP.

In the course of induced scattering, the intensity of the scattered component grows exponentially as long as it is less than the original MP intensity. At this stage the intensities are correlated. In the framework of our model, the recently discovered stiff connection of the subpulse modulation in the IP and MP of PSR B1702-19 \citep{p2-w07} is explained naturally. 

As soon as the scattered component gains a substantial part of the MP energy, its growth enters the stage of saturation and the MP intensity is markedly suppressed. This is believed to take place in PSR J1326-6700, where the PR is seen only when the MP is below the detection level \citep*{p2-wang}.

\subsection{Population statistics}

The efficiency of induced scattering depends on the basic pulsar parameters and on the characteristics of the plasma flow in the scattering region.	Strong enough scattered components are expected in the pulsars with	{\itshape i)} short periods, {\itshape ii)} strong magnetic fields, and {\itshape iii)}	high radio luminosities.
At the stage of exponential growth, the scattered intensity may strongly fluctuate as a result of small variations in the plasma flow, so that	the components with the intensities below the detection level may occasionally become observable. This implies the possibility of transient components outside of the MP.

\section{Conclusions}

We for the first time suggest the physical mechanism of the component formation outside of the MP. In our model, the PR and IP result from the induced scattering of the MP radiation into the background. This mechanism implies the component connection at widely spaced frequencies.	Our model explains
{\itshape i)}	the component location on the pulse profile;
{\itshape ii)} the relative strengthening of the scattered components at low frequencies;
{\itshape iii)}	the complete linear polarization of the PR;
{\itshape iv)}	the high linear polarization of the IP;
{\itshape v)}	the constancy of the position angle across the scattered components; and
{\itshape vi)}	the correlation and anticorrelation of the components with the MP and with each other. We also predict the possibility of transient components outside of the MP.

\acknowledgements I am grateful to the LFRU organizers for the financial assistance which has made possible my participation in the Conference. The work is partially supported by INTAS Grant No 03-5727.

\end{document}